\documentclass[letterpaper]{jpconf}
\usepackage{graphicx}
\usepackage{float}
\usepackage{citesort}

\usepackage{subcaption}
\usepackage{amssymb}
\usepackage{amsmath}

\usepackage{ulem}

\newcommand{\thornado}{\texttt{thornado}}
\newcommand{\amrex}{\texttt{AMReX}}

\newcommand{\Msun}{M_{\odot}}

\usepackage{xcolor}

\usepackage{soul} 
\setstcolor{blue}

\renewcommand{\bs}[1]{\boldsymbol{#1}}
\newcommand{\mc}[1]{\mathcal{#1}}

\newcommand{\figref}[1]{Figure~\ref{#1}}
\newcommand{\secref}[1]{Section~\ref{#1}}

\begin{document}
\nocite{*}

\title{A discontinuous Galerkin method for general relativistic hydrodynamics in \thornado}

\author{Samuel J Dunham\textsuperscript{1,2}, E Endeve\textsuperscript{3,2}, A Mezzacappa\textsuperscript{2,4}, J Buffaloe\textsuperscript{2}, and K Holley-Bockelmann\textsuperscript{1,5}}

\address{\textsuperscript{1} Department of Astronomy, Vanderbilt University, 6301 Stevenson Center Lane, Nashville TN, 37235, USA}
\address{\textsuperscript{2} Department of Physics and Astronomy, University of Tennessee-Knoxville, Nielsen Physics Building, 401, 1408 Circle Drive, Knoxville TN, 37996, USA}
\address{\textsuperscript{3} Computer Science and Mathematics Division, Oak Ridge National Laboratory, Oak Ridge, TN 37831, USA}
\address{\textsuperscript{4} Joint Institute for Computational Sciences, Oak Ridge National Laboratory, Oak Ridge, TN 37831, USA}
\address{\textsuperscript{5} Department of Life and Physical Sciences, Fisk University, 1000 17\textsuperscript{th} Ave N, Nashville TN, 37208, USA}

\ead{samuel.j.dunham@vanderbilt.edu}

\begin{abstract}
Discontinuous Galerkin (DG) methods provide a means to obtain high-order accurate solutions in regions of smooth fluid flow while, with the aid of limiters, still resolving strong shocks. These and other properties make DG methods attractive for solving problems involving hydrodynamics; e.g., the core-collapse supernova problem. With that in mind we are developing a DG solver for the general relativistic, ideal hydrodynamics equations under a 3+1 decomposition of spacetime, assuming a conformally-flat approximation to general relativity. With the aid of limiters we verify the accuracy and robustness of our code with several difficult test-problems: a special relativistic Kelvin--Helmholtz instability problem, a two-dimensional special relativistic Riemann problem, and a one- and two-dimensional general relativistic standing accretion shock (SAS) problem. We find good agreement with published results, where available. We also establish sufficient resolution for the 1D SAS problem and find encouraging results regarding the standing accretion shock instability (SASI) in 2D.

\end{abstract}

\section{Introduction}
Core-collapse supernovae are multi-physics, multi-dimensional phenomena that require sophisticated numerical methods to accurately capture all of their features, both on a macroscopic and a microscopic scale. A detailed explanation of the physical processes involved is beyond the scope of this document, but here we present a brief overview, closely following \cite{Mezzacappa2005}. When the core of a massive star ($M\gtrsim10 M_{\odot}$) reaches the Chandrasekhar mass-limit, electron degeneracy pressure can no longer support the core against gravity, and collapse ensues. As the iron-core collapses, it splits into a subsonically collapsing inner-core and a supersonically collapsing outer-core; the inner-core continues to collapse until nuclear densities are reached, at which point it undergoes a phase transition from a heterogeneous ensemble of distinct nuclei and nucleons to bulk nuclear matter. This transition causes the equation of state to stiffen and the inner-core to ``bounce", generating a shock wave. The shock wave propagates outward, losing energy to the dissociation of nuclei and to neutrino emission, until it stalls at a distance $\mc{O}$(100 km) from the center. Understanding how the shock wave is reenergized is one of the goals of modern core-collapse supernova (CCSN) science.

Broadly speaking there are three branches of physics that every realistic CCSN model must treat faithfully: gravity, neutrino transport, and hydrodynamics. To model these events we are developing the \textbf{t}oolkit for \textbf{h}igh-\textbf{or}der \textbf{n}eutrino-r\textbf{ad}iation hydr\textbf{o}dynamics---\thornado. \thornado\ is a software package designed to solve the equations of neutrino transport (using a two-moment method \cite{Cardall2013,Shibata2011}) and hydrodynamics using a discontinuous Galerkin (DG) method \cite{CockburnShu1998,HesthavenWarburtonNodalDGMethods}. It has been partially coupled to \amrex\ \cite{AMReX}, a framework for adaptive mesh refinement and distributed parallel computing. This document focuses on the solver for the hydrodynamics equations, which employs a DG method. The events we will ultimately be modeling will involve general relativistic conditions, and therefore we begin with the general relativistic (GR) Euler equations of hydrodynamics (zero physical viscosity) in curvilinear coordinates. The neglect of physical viscosity is a valid approximation since the Reynolds numbers are sufficiently large that energy dissipation occurs on scales much smaller than those in which we are interested \cite{ThompsonDuncan1993}. A recent review on the requirements for modeling the hydrodynamics in CCSNe can be found in \cite{Muller2016}. The equations we solve assume a 3+1 decomposition of spacetime as well as the conformally-flat approximation, which is acceptable for slowly-rotating progenitors \cite{Dimmelmeier2002}.

There are a number of CCSN codes currently in use \cite{AGILEBOLTZTRAN,CHIMERA,FORNAX,ALCAR,PROMETHEUSVERTEX,COCONUTVERTEX,ZELMANI,FLASH,GR1D,Kotake,Nagakura}, each handling the hydrodynamics in its own way and offering varying orders of accuracy; however, they all use different versions of finite-volume and finite-difference methods. Here we use the DG method, a finite-element method, with advantages such as high-order accuracy on a compact stencil and $hp$-adaptivity. In part because of the advantages offered by DG methods, they have found application to this and other branches of astrophysics; e.g., radiative transfer \cite{Chu2019,Endeve2015,Radice2013,Kitzmann2016}, numerical relativity \cite{Teukolsky2016}, and turbulence \cite{BauerDG}.

\section{Physical Model}
We solve the 3+1 GR hydrodynamics (GRHD) equations in the Valencia formulation, assuming for the moment a stationary spacetime and vanishing shift vector, $\beta^{i}=0$. Using units where the speed of light $c=1$, the equations then take the form \cite{RezzollaZanottiRelativisticHydrodynamics}
\begin{equation}\label{Eq:Hydro}
  \sqrt{\gamma}\,\partial_{t}\,\bs{U}+\partial_{i}\left(\alpha\,\sqrt{\gamma}\,\bs{F}^{i}\left(\bs{U}\right)\right)=\alpha\,\sqrt{\gamma}\,\bs{S}\left(\bs{U}\right),\hspace{1em}\bs{U}=\bs{U}\left(\bs{x},t\right),\hspace{1em}\bs{x}\in\mathbb{R}^{3},\hspace{1em}t\in\mathbb{R},
\end{equation}
where $\bs{U}$ is the vector of conserved variables, $\bs{F}^{i}$ is the vector of fluxes, with $i$ specifying the spatial dimension ($i=1,\cdots,d$, where $d$ is the spatial dimensionality of the problem considered), $\bs{S}$ is a source term, $\alpha$ is the lapse function (measuring elapsed proper time between adjacent spatial hypersurfaces), and $\sqrt{\gamma}$ is the square root of the determinant of the spatial three-metric, $\gamma_{ij}$. The conserved variables are $\bs{U}=\left(D,\ S_{j},\ \tau\right)^{\top}$, where $D$ is the conserved rest-mass density, $S_{j}$ is the component of the conserved momentum density in the $j$\textsuperscript{th} spatial dimension ($j=1,\cdots,d$), and $\tau$ is the conserved energy density with the conserved rest-mass density extracted (the superscript $\top$ means transpose). These quantities are related to the primitive variables $\bs{V}=\left(\rho,\ v^{j},\ e\right)^{\top}$, where $\rho$ is the rest-mass density, $v^{j}$ is the component of the fluid three-velocity in the $j$\textsuperscript{th} spatial dimension, and $e$ is the internal energy density, assumed in this document to be related to the pressure $p$ by an ideal equation of state: $p=\left(\Gamma-1\right)e$, where $\Gamma\in\left(1,2\right]$ is the ratio of specific heats. The relations between the conserved and primitive variables are
 \begin{equation}\label{Eq:Prim2Cons}
     \bs{U}\left(\bs{V}\right)=\begin{pmatrix}D\\ S_{j}\\\tau\end{pmatrix}=\begin{pmatrix}\rho\,W\\ \rho\,h\,W^{2}\,v_{j}\\\rho\,h\,W^{2}-p-\rho\,W\end{pmatrix},
 \end{equation}
 where $W=\left(1-v^{i}v_{i}\right)^{-1/2}$ is the Lorentz factor and $h=1+\left(e+p\right)/\rho$ is the relativistic specific enthalpy. The fluxes are given by
 \begin{equation}
     \bs{F}^{i}\left(\bs{U}\right)=\begin{pmatrix}D\,v^{i}\\ P^{i}_{~j}\\ S^{i}-D\,v^{i}\end{pmatrix},
 \end{equation}
 where the $P^{i}_{~j}$ are components of the pressure tensor, defined as $P^{ij}=\rho\,h\,W^{2}\,v^{i}\,v^{j}+p\,\gamma^{ij}$, where $\gamma^{ij}$ is the inverse of $\gamma_{ij}$; i.e., $\gamma^{ik}\gamma_{kj}=\delta^{i}_{~j}$. The sources are given by
 \begin{equation}
     \bs{S}\left(\bs{U}\right)=\begin{pmatrix}0\\\frac{1}{2}\,P^{ik}\,\partial_{j}\,\gamma_{ik}-\alpha^{-1}\left(\tau+D\right)\,\partial_{j}\,\alpha\\-\alpha^{-1}\,S^{j}\,\partial_{j}\,\alpha\end{pmatrix}.
 \end{equation}

 \section{Numerical Method}

 \subsection{Overview of the DG method}
 The DG method provides a means of solving partial differential equations and obtaining high-order accurate solutions in space, on a compact stencil. This is achieved by using a high-degree polynomial representation of the solution within each element, communicating with nearest-neighbors only. This makes the DG method well-suited for massively parallel architectures, which is a highly sought-after property for, among other purposes, high-resolution 3D CCSN simulations. We give a brief overview of the DG method here.

 \subsubsection{Basis Functions}
 We start by discretizing the spatial domain into a set of $\mc{N}_{E}$ non-overlapping elements $\bs{K}$, where
 \begin{equation}
     \bs{K}=\left\{\bs{x}:x^{i}\in K^{i}\equiv\left(x^{i}_{L},x^{i}_{U}\right),i=1,\cdots,d\right\},
 \end{equation}
 where $x^{i}_{L(U)}$ are the lower(upper) boundaries of $x^{i}$ in the element $\bs{K}$. We define an approximation space, $\mathbb{V}^{k}$, chosen to be the space spanned by the $d$-dimensional tensor product of 1D Lagrange polynomials with degree less than or equal to $k$. We then seek to approximate the exact solution $\bs{U}$ in each element $\bs{K}$ with $\bs{U}_{h}\in\mathbb{V}^{k}$.
 For the case $d=3$ this takes the form
 \begin{equation}\label{Eq:ApproxSolution}
     \bs{U}\left(\bs{x},t\right)\approx\bs{U}_{h}\left(\bs{x},t\right)=\sum\limits_{i_{3}=1}^{N}\sum\limits_{i_{2}=1}^{N}\sum\limits_{i_{1}=1}^{N}\bs{U}_{h}\left(x^{1}_{i_{1}},x^{2}_{i_{2}},x^{3}_{i_{3}},t\right)\,\ell_{i_{1}}\left(x^{1}\right)\,\ell_{i_{2}}\left(x^{2}\right)\,\ell_{i_{3}}\left(x^{3}\right),\hspace{1em}x^{i}\in K^{i},
 \end{equation}
 where $N=k+1$ is the chosen number of interpolation points per element, per spatial dimension, and $\ell_{i_{1}}\left(x^{1}\right)$ is the $i_{1}$\textsuperscript{th} Lagrange polynomial in the $x^{1}$-direction on the $\bs{K}$\textsuperscript{th} element, defined as
 \begin{equation}
     \ell_{i_{1}}\left(x^{1}\right)=\prod\limits_{\substack{j_{1}=1 \\ j_{1}\neq i_{1}}}^{N}\frac{x^{1}-x^{1}_{j_{1}}}{x^{1}_{i_{1}}-x^{1}_{j_{1}}},\hspace{1em}i_{1}=1,\cdots,N,
 \end{equation}
 with the other Lagrange polynomials defined similarly. It is understood that the Lagrange polynomials have compact support in $\bs{K}$. Note that $\ell_{i_{1}}\left(x^{1}_{j_{1}}\right)=\delta_{i_{1}j_{1}}$, so that interpolation to a grid point within $\bs{K}$ yields the value of $\bs{U}_{h}$ at that grid point.
In \thornado\ we choose to have the grid points coincide with the Gauss--Legendre quadrature points because this simplifies the implementation and reduces the amount of computational work required. To simplify the notation we introduce a multi-index, $\bs{i}\equiv\left\{i_{1},\cdots,i_{d}\right\}$, and define $\bs{x}_{\bs{i}}\equiv\left\{x^{1}_{i_{1}},\cdots,x^{d}_{i_{d}}\right\}$, so that the approximation \eqref{Eq:ApproxSolution} can be written as
 \begin{equation}\label{Eq:Approx}
     \bs{U}_{h}\left(\bs{x},t\right)=\sum\limits_{\bs{i}=\bs{1}}^{\bs{N}}\bs{U}_{\bs{i}}\left(t\right)\,\ell_{\bs{i}}\left(\bs{x}\right),
 \end{equation}
 where $\bs{U}_{\bs{i}}\left(t\right)\equiv\bs{U}_{h}\left(\bs{x}_{\bs{i}},t\right)$, and $\ell_{\bs{i}}\left(\bs{x}\right)=\prod_{j=1}^{d}\ell_{i_{j}}\left(x^{j}\right)$. Note that the delta-function property carries over; i.e., $\ell_{\bs{i}}\left(\bs{x}_{\bs{j}}\right)=\delta_{\bs{ij}}$.


 \subsubsection{The Galerkin Method}
 To obtain a numerical scheme for solving \eqref{Eq:Hydro} we first convert \eqref{Eq:Hydro} into a system of ordinary differential equations, which we solve with the Galerkin method. To start, we multiply \eqref{Eq:Hydro} by a function in our approximation space $\mathbb{V}^{k}$, substitute our approximate solution, \eqref{Eq:Approx}, for $\bs{U}$, integrate over the volume of the element $\bs{K}$, and perform integration-by-parts on the flux term, which gives
 \begin{align}\label{Eq:WeakForm1}
     \int_{\bs{K}}\partial_{t}\left(\bs{U}_{h}\right)\,\ell_{\bs{j}}\,dV&+\sum\limits_{i=1}^{d}\int_{\partial\bs{K}^{i}}\left[\alpha\,\sqrt{\gamma}\,\ell_{\bs{j}}\,\widehat{\bs{F}^{i}}\left(\bs{U}_{h}\right)\right]^{x^{i}=x^{i}_{U}}_{x^{i}=x^{i}_{L}}d\tilde{\bs{x}}^{i}&\notag\\
     &-\sum\limits_{i=1}^{d}\int_{\bs{K}}\alpha\,\bs{F}^{i}\left(\bs{U}_{h}\right)\,\partial_{i}\,\ell_{\bs{j}}\,dV=\int_{\bs{K}}\alpha\,\ell_{\bs{j}}\,\bs{S}\left(\bs{U}_{h}\right)\,dV,\hspace{1em}\bs{j}=\bs{1},\cdots,\bs{N},
 \end{align}
where $d$ is the dimensionality of the problem, $\partial\boldsymbol{K}^{i}$ represents the element interface in the $x^{i}$-direction, $\tilde{\bs{x}}^{i}$ refers to the coordinates that are \textit{not} $x^{i}$ [e.g., if $d=3$ and $x^{i}=x^{1}$, then $\tilde{\bs{x}}^{i}=\left(x^{2},x^{3}\right)$], $dV=\sqrt{\gamma}\prod_{i=1}^{d}dx^{i}$, and we have substituted the numerical flux $\widehat{\bs{F}^{i}}$ (which we obtain with an approximate Riemann solver) for the flux evaluated at the element interfaces.

Performing the integrals in \eqref{Eq:WeakForm1} with the $N$-point Gauss--Legendre quadrature yields a modified version of \eqref{Eq:WeakForm1} which we solve for $\bs{U}_{\bs{j}}$:
\begin{align}\label{Eq:ODE}
  \frac{d\bs{U}_{\bs{j}}}{dt}&=-\frac{1}{\sqrt{\gamma_{\bs{j}}}}\sum\limits_{i=1}^{d}\frac{1}{w_{j_{i}}\,\Delta x^{i}}\left[\alpha\left(x^{i},\tilde{\bs{x}}^{i}_{\tilde{\bs{j}}_{i}}\right)\,\sqrt{\gamma\left(x^{i},\tilde{\bs{x}}^{i}_{\tilde{\bs{j}}_{i}}\right)}\,\widehat{\bs{F}^{i}}\left(x^{i},\tilde{\bs{x}}^{i}_{\tilde{\bs{j}_{i}}},t\right)\,\ell_{j_{i}}\left(x^{i}\right)\right]^{x^{i}=x^{i}_{U}}_{x^{i}=x^{i}_{L}}\notag\\
  &\hspace{1em}+\frac{1}{\sqrt{\gamma_{\bs{j}}}}\sum\limits_{i=1}^{d}\frac{1}{w_{j_{i}}\,\Delta x^{i}}\sum\limits_{q_{k}=1}^{N}\Bigg[w_{q_{k}}\,\alpha\left(x^{i}_{q_{k}},\tilde{\bs{x}}^{i}_{\tilde{\bs{j}}_{i}}\right)\,\sqrt{\gamma\left(x^{i}_{q_{k}},\tilde{\bs{x}}^{i}_{\tilde{\bs{j}}_{i}}\right)}\notag\\
  &\hspace{10em}\times\bs{F}^{i}\left(\bs{U}_{h}\left(x^{i}_{q_{k}},\tilde{\bs{x}}^{i}_{\tilde{\bs{j}}_{i}},t\right)\right)\,\partial_{i}\,\ell_{j_{i}}\left(x^{i}_{q_{k}}\right)\Bigg]+\alpha_{\bs{j}}\,\bs{S}_{\bs{j}},\hspace{1em}\bs{j}=\bs{1},\cdots,\bs{N}.
\end{align}
The scalars $w_{j_{i}}$ are the Gauss--Legendre quadrature weights and $\sqrt{\gamma_{\bs{j}}}=\sqrt{\gamma\left(\bs{x_{\bs{j}}}\right)}$, $\alpha_{\bs{j}}=\alpha\left(\bs{x}_{\bs{j}}\right)$, etc. Note that due to our choice of the collocation method we are able to exploit the delta-function properties of the Lagrange polynomials. This method is similar to that used by \cite{Bassi2013}. With \eqref{Eq:ODE} we now have a set of ordinary differential equations (ODEs) we can evolve in time with an ODE integrator \cite{CockburnShu2001}.

\subsection{Time-Stepping}
We integrate the system of ODEs, \eqref{Eq:ODE}, with explicit strong-stability-preserving Runge-Kutta methods \cite{ShuOsher1988}. The methods use convex combinations of forward-Euler time steps and therefore inherit the numerical stability properties of the forward-Euler method: Subject to a CFL restriction on the time step, we will have stability of the cell averages and also high-order accuracy in time. Denoting by $u^{n}$ the solution for all of the variables across the entire spatial domain at a time $t^{n}$, the algorithm is given by
 \begin{alignat}{2}
 1.)&\hspace{0.5em}u^{\left(0\right)}&&=u^{n},\notag\\
 2.)&\hspace{0.5em}u^{\left(i\right)}&&=\Lambda_{\text{BP}}\left[\Lambda_{\text{TVD}}\left[u^{\left(0\right)}+\sum\limits_{j=0}^{i-1}c_{ij}\,\Delta t^{n}\,L\left(u^{\left(j\right)}\right)\right]\right],\notag\\
 &\hspace{1em}&&=\Lambda_{\text{BP}}\left[\Lambda_{\text{TVD}}\left[\sum\limits_{j=0}^{i-1}\alpha_{ij}\left\{u^{\left(j\right)}+\frac{\beta_{ij}}{\alpha_{ij}}\,\Delta t^{n}\,L\left(u^{\left(j\right)}\right)\right\}\right]\right],\hspace{1em}i=1,\cdots,N_{s},\notag\\
 3.)&\hspace{0.5em}u^{n+1}&&=u^{\left(N_{s}\right)},\label{Eq:SSP}
 \end{alignat}
where $c_{ij}$ (and, equivalently, $\alpha_{ij}$ and $\beta_{ij}$) are constants that satisfy certain conditions that guarantee a specified order of accuracy. From the second line of step 2 in \eqref{Eq:SSP} we identify the intermediate stages as convex combinations of forward-Euler steps with time step $(\beta_{ij}/\alpha_{ij})\Delta t^{n}$, and therefore the $\alpha_{ij}$ must all be non-negative and satisfy the relation $\sum_{j}\alpha_{ij}=1,\forall i=1,\cdots,N_{s}$. Further, for these to be forward-Euler steps with positive time step, all of the $\beta_{ij}$ must be positive.

Note that for $N_{s}=1$ (in which case $\alpha_{10}=\beta_{10}=c_{10}=1$) we recover the simple forward-Euler scheme. All of the results shown in this document were run with $N_{s}=3$, the highest available order for this method. The non-linear operators $\Lambda_{\text{TVD}}$ and $\Lambda_{\text{BP}}$ are respectively the slope and bound-preserving limiters, which are discussed in following sections. To determine the time step, we use an estimate of the largest wave-speed, $\left|\lambda^{i}\right|$, over the entire spatial domain, in conjunction with the formula given in \cite{CockburnShu2001}:
\begin{equation}
  \Delta t=\frac{C_{\text{CFL}}}{d\left(2k+1\right)}\text{min}\left[\frac{\Delta x^{1}}{\left|\lambda^{1}\right|},\cdots,\frac{\Delta x^{d}}{\left|\lambda^{d}\right|}\right],
\end{equation}
where $k$ is the degree of the polynomial approximation we use, $C_{\text{CFL}}$ is a dimensionless number of order unity, and $d$ is the spatial dimensionality of the problem.

\subsection{Slope Limiter}
Due to the polynomial approximation of the solution, spurious oscillations can develop near strong gradients (e.g., shocks). The method we use to mitigate these oscillations follows \cite{CockburnShu1998}, in which the local solution is projected into a space of orthogonal polynomials with a hierarchical structure (e.g., Legendre polynomials), and if limiting is deemed necessary the solution is truncated after the linear term, which itself is modified. Therefore, in regions where limiting has been applied, our solution is at most 2\textsuperscript{nd}-order accurate. For example, the expansion coefficient for the linear term in the $x^{1}$-dimension, $c_{1}$, is modified according to the MinMod limiter (ignoring constant normalization factors):
\begin{equation}
\widetilde{c}_{1}=\text{MinMod}\left(c_{1},\beta_{\text{TVD}}\frac{\bs{U}_{\bs{K}}-\bs{U}_{K^{1}-1}}{\Delta x^{1}},\beta_{\text{TVD}}\frac{\bs{U}_{K^{1}+1}-\bs{U}_{\bs{K}}}{\Delta x^{1}}\right),
\end{equation}
where the MinMod limiter is defined as in \cite{CockburnShu1998}. $\bs{U}_{\bs{K}}$ is the cell average of the solution in the $\bs{K}$\textsuperscript{th} element and, $\bs{U}_{K^{1}\pm1}$ are the cell averages of the neighbors in the $x^{1}$-direction. The parameter $\beta_{\text{TVD}}$ affects the severity of the limiting; it takes values between 1 and 2, and optimal values seem to be problem-dependent. Further investigation is required to determine the ideal value for the CCSN problem. To ensure that limiting is only applied in the vicinity of strong discontinuities, we also implement a shock-detector, or a \textit{troubled-cell indicator}, for which we use the method discussed in \cite{FuShu2017}. We also find that superior results are obtained when limiting the characteristic variables \cite{CockburnShu2001} instead of the individual (coupled) components (see \secref{sec.results}).


\subsubsection{Conservative Correction}
Our code is generalized for curvilinear coordinates, and this poses an issue when it comes to ensuring conservation, particularly in the slope-limiting procedure, because we approximate $\bs{U}_{h}$ and not $\sqrt{\gamma}\,\bs{U}_{h}$. The modal form, mentioned above, before and after limiting, looks like
\begin{equation}\label{eq.minmod}
\bs{U}_{h}\left(\bs{x},t\right)=\sum\limits_{\bs{n}=\bs{0}}^{\bs{N}-\bs{1}}c_{\bs{n}}\left(t\right)P_{\bs{n}}\left(\bs{x}\right)\longrightarrow\widetilde{\bs{U}}_{h}\left(\bs{x},t\right)=\sum\limits_{\bs{n}=\bs{0}}^{\bs{1}}\widetilde{c}_{\bs{n}}\left(t\right)P_{\bs{n}}\left(\bs{x}\right),
\end{equation}
where $P_{\bs{n}}\left(\bs{x}\right)$ is a $d$-dimensional tensor product of Legendre polynomials, and $c_{\bs{0}}$ corresponds to the cell average. In curvilinear coordinates, the cell average is modified by the MinMod limiter, i.e., when $c_{\bs{0}}\longrightarrow\widetilde{c}_{\bs{0}}$, thus destroying the conservative aspect of the algorithm. To recover conservation, we apply an a posteriori \textit{conservative correction} to the elements that were limited. The correction is applied only to the cell average, and is computed by substituting \eqref{eq.minmod} into the formula for the cell average, and setting that equal to $c_{\bs{0}}$. By doing so, the value of the cell average after limiting is identical to its value before limiting (see also \cite{RadiceRezzolla2011}):
\begin{equation}
\widetilde{c}_{\bs{0}}=c_{\bs{0}}-\widetilde{c}_{\bs{1}}\frac{\int_{\bs{K}}P_{\bs{1}}\left(\bs{x}\right)\sqrt{\gamma}d^{3}x}{\int_{\bs{K}}\sqrt{\gamma}d^{3}x},
\end{equation}
where there is an implied summation over repeated multi-index, with the sum going over the active spatial dimensions of the problem.

%

\subsection{Bound-Preserving Limiter}
Physically it is true that $p\geq0$, $\rho\geq0$ and $\left|v\right|<c$, but numerically these conditions can be violated, especially in problems involving strong shocks. However, due to the restriction we impose on the time step, the cell average at time $t^{n+1}$ will always be physical\footnote{This is not strictly true in the case of curvilinear coordinates; however in practice we find that it still holds.}, provided that the solution is physical in a discrete number of quadrature points within each element at $t^{n}$. However, it is possible that the solution at time $t^{n+1}$ is unphysical in some of the quadrature points. The method we use to mitigate this is given in \cite{Qin2016}: The idea is to form a convex combination of the cell average and the unphysical points, and then damp those unphysical points towards the cell average until a physically valid solution is obtained. This is possible because the set of points that are physically admissible is convex.

With all the pieces in place we are ready to show some preliminary results obtained with our solver.

\section{Preliminary Numerical Results}\label{sec.results}
We have benchmarked our code with several challenging test problems, some of which are highlighted below. For all of these tests we use third-order methods in space and time.

\subsection{Relativistic Kelvin--Helmholtz Instability Problem}
\begin{figure}
\centering
\includegraphics[width=0.8\textwidth]{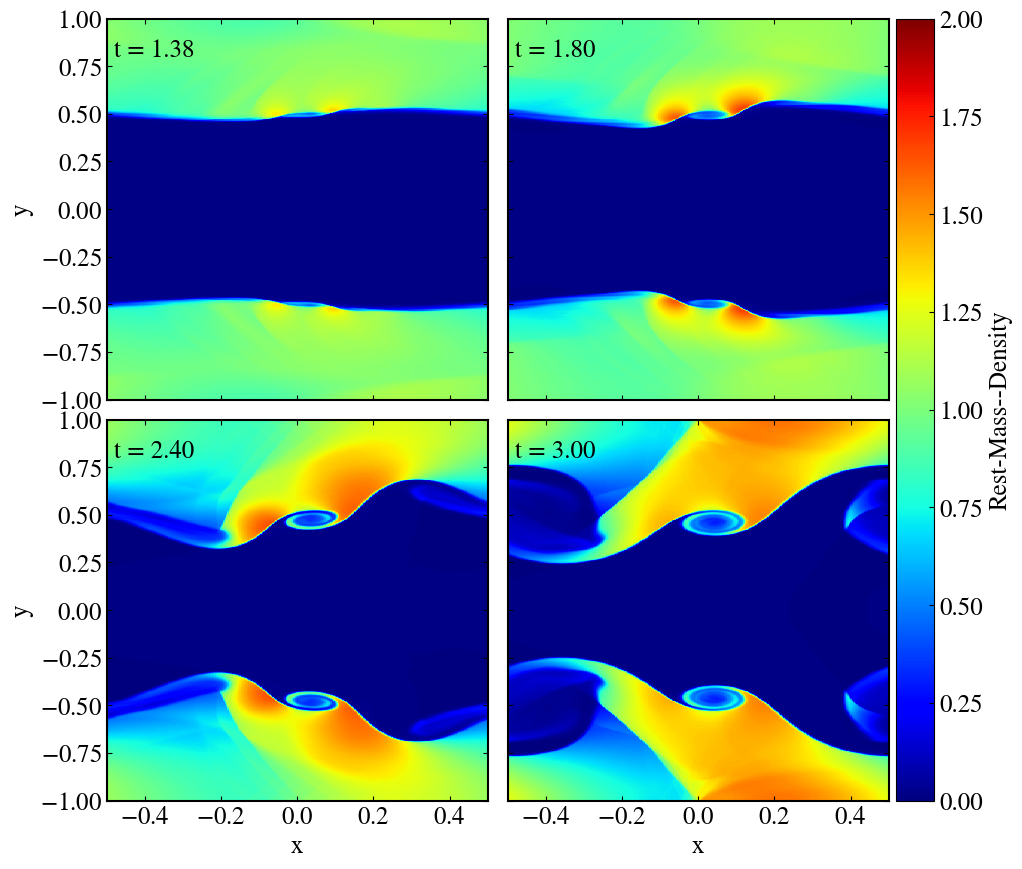}
\caption{Rest-mass density for the relativistic Kelvin--Helmholtz instability problem at times $t=1.38$ (top-left), $t=1.80$ (top-right), $t=2.40$ (bottom-left), and $t=3.00$ (bottom-right). The instability can be seen to grow and develop from the linear to the non-linear regime, and the final results agree qualitatively with published results in \cite{RadiceRezzolla2012}.}
\label{Fig:KHI}
\end{figure}

Our first test is a special relativistic Kelvin--Helmholtz instability problem, a ubiquitous problem in astrophysics, from \cite{RadiceRezzolla2012}. This problem has been used to verify accuracy by comparing growth rates to predictions from linear theory, but this is not our focus. The setup includes a velocity shear in the $y$-direction, an adiabatic index $\Gamma=4/3$, and a computational domain extending from $x\in\left[-0.5,+0.5\right]$ and $y\in\left[-1,+1\right]$, with periodic boundary conditions in both dimensions. This problem tests the code's ability to resolve turbulent regions in smooth fluid flow. The shear velocity profile is given by
\begin{equation}
v^{x}\left(x,y\right)=\begin{cases}+V_{\text{shear}}\,\text{tanh}\left[\left(y-0.5\right)/a\right],&\text{if }y>0\\-V_{\text{shear}}\,\text{tanh}\left[\left(y+0.5\right)/a\right],&\text{if }y\leq0\end{cases},
\end{equation}
where $V_{\text{shear}}=0.5$, and $a=0.01$.
The instability is introduced with a single-mode velocity perturbation given by
\begin{equation}
v^{y}\left(x,y\right)=\begin{cases}+A_{0}\,V_{\text{shear}}\,\sin\left(2\pi x\right)\,\exp\left[-\left(y-0.5\right)^{2}/\sigma^{2}\right],&\text{if }y>0\\-A_{0}\,V_{\text{shear}}\,\sin\left(2\pi x\right)\,\exp\left[-\left(y+0.5\right)^{2}/\sigma^{2}\right],&\text{if }y\leq0\end{cases},
\end{equation}
where $A_{0}=0.1$, and $\sigma=0.1$. The rest-mass density is given by
\begin{equation}
\rho\left(y\right)=\begin{cases}\rho_{0}+\rho_{1}\,\text{tanh}\left[\left(y-0.5\right)/a\right],&\text{if }y>0\\\rho_{0}-\rho_{1}\,\text{tanh}\left[\left(y+0.5\right)/a\right],&\text{if }y\leq0\end{cases},
\end{equation}
where $\rho_{0}=0.505$ and $\rho_{1}=0.495$. For this problem we used the HLLC Riemann solver \cite{MignoneBodo2005}. In \figref{Fig:KHI} we show the rest-mass density from a simulation run with a resolution of $256\times512$ and $C_{\text{CFL}}=0.5$.

\subsubsection{Strong-Scaling Test}
\begin{figure}
  \centering
  \includegraphics[width=0.9\textwidth]{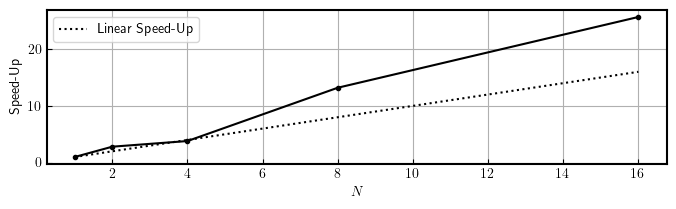}
  \caption{Strong-scaling results for a 3D Kelvin--Helmholtz instability problem with $32^{3}$ elements using 1, 2, 4, 8, and 16 MPI processes, normalized to wall-time when running serially. The dotted line shows what we would expect from linear scaling. We consistently achieve at least linear scaling.}
  \label{Fig:StrongScaling}
\end{figure}

We have begun to interface the distributed parallelism capabilities of \amrex\ into \thornado, and have performed a strong-scaling test using the 3D Kelvin--Helmholtz instability problem from \cite{RadiceRezzolla2012}. Using $32^3$ elements we ran with 1, 2, 4, 8, and 16 processes, and found good scaling, as shown in \figref{Fig:StrongScaling}. Future work includes extending this to more processes, as well as performing weak-scaling tests.

\subsection{Relativistic 2D Riemann Problem}
\begin{figure}
  \centering
  \includegraphics[width=0.8\textwidth]{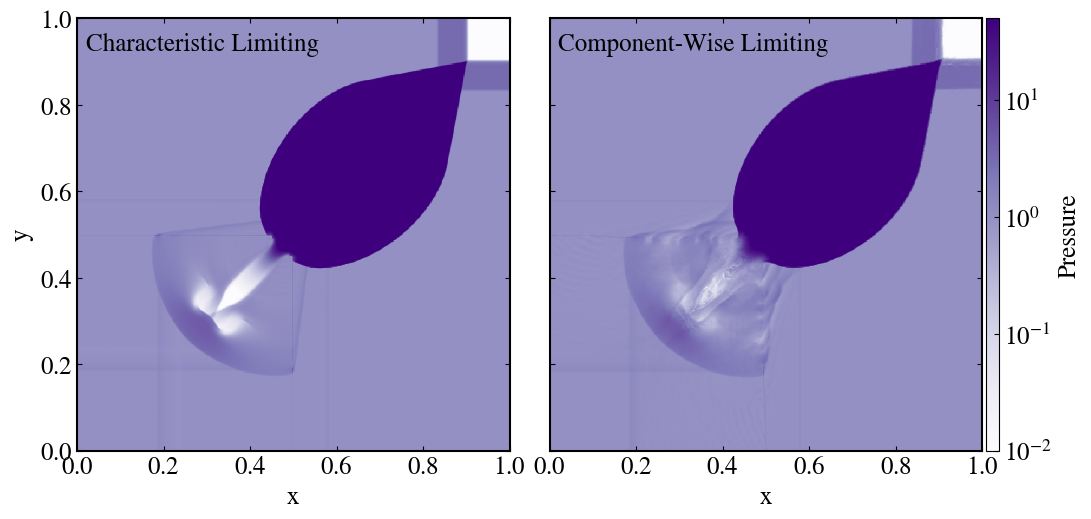}
   \caption{Log plot of the pressure at $t=0.4$ for a relativistic 2D Riemann problem from \cite{DelZannaBucciantini2002}. The left panel shows the results obtained using characteristic limiting and the right plot shows the results obtained using component-wise limiting. The pressure in the right plot is much more oscillatory, particularly in the lower left quadrant, showing the advantages of using characteristic limiting.}
   \label{Fig:2DRP}
 \end{figure}

The next test is a special relativistic, 2D Riemann problem from \cite{DelZannaBucciantini2002}. This problem tests the code's ability to resolve contact discontinuities and strong shocks in multiple dimensions, the initial Lorentz factor being $W\sim7$. The boundary conditions allow for free expansion in both directions, and the computational domain of $x,y\in\left[0,1\right]$ with a resolution of $256\times256$ is initially divided into four quadrants. The adiabatic index $\Gamma=5/3$ and the initial conditions are
 \begin{equation}
 \rho=\begin{cases}0.1,&\text{NE}\\0.1,&\text{NW}\\0.5,&\text{SW}\\0.1,&\text{SE}\end{cases},\hspace{1em}v^{x}=\begin{cases}0.0,&\text{NE}\\0.99,&\text{NW}\\0.0,&\text{SW}\\0.0,&\text{SE}\end{cases},\hspace{1em}v^{y}=\begin{cases}0.0,&\text{NE}\\0.0,&\text{NW}\\0.0,&\text{SW}\\0.99,&\text{SE}\end{cases},\hspace{1em}p=\begin{cases}0.01,&\text{NE}\\0.1,&\text{NW}\\0.1,&\text{SW}\\0.1,&\text{SE}\end{cases}.
 \end{equation}
The simulation is run until $t=0.4$ with the HLLC Riemann solver and $C_{\text{CFL}}=0.5$, and the final pressure is shown in \figref{Fig:2DRP}.

 \subsection{Relativistic 1D SAS Problem}
Next we show results from our first general-relativistic test: A one-dimensional, spherically symmetric, perturbed, standing accretion shock (SAS). This problem tests the code's implementation of curvilinear coordinates as well as the GR aspect of the GRHD equations. We assume a stationary background spacetime given by the Schwarzschild metric in isotropic coordinates (in units where $c=G=1$) \cite{BaumgarteShapiroNumericalRelativity}:
 \begin{align}
 ds^{2}&=-\alpha^{2}\,dt^{2}+\psi^{4}\left(dr^{2}+\,r^{2}\,d\Omega^{2}\right)\notag\\
 &=-\left(\frac{1-M/2r}{1+M/2r}\right)^{2}dt^{2}+\left(1+\frac{M}{2r}\right)^{4}\left[dr^{2}+r^{2}\left(d\theta^{2}+\sin^{2}\theta\,d\varphi^{2}\right)\right]\\
 &\approx-\left(1+2\,\Phi\right)dt^{2}+\left(1-2\,\Phi\right)\left[dr^{2}+r^{2}\left(d\theta^{2}+\sin^{2}\theta\,d\varphi^{2}\right)\right],\label{eq.NewtonianLapse}
 \end{align}
 where $\alpha\equiv\left|\left(1-M/2r\right)/\left(1+M/2r\right)\right|$ is the lapse function, $\psi\equiv\left|1+M/2r\right|$ is the conformal factor, and we choose a mass (corresponding to the mass of the proto-neutron star, or PNS) $M$ of $1.4\,\Msun$. The third line is an approximation, valid for $\Phi\ll1$, where $\Phi=-M/r$ is the Newtonian gravitational potential. In this limit, the lapse function is $\alpha_{N}=\left(1+2\,\Phi\right)^{1/2}$ and the conformal factor is $\psi_{N}=\left(1-2\,\Phi\right)^{1/4}$. To determine the initial conditions we solve the GRHD equations assuming a steady-state (no explicit time-dependence), spherical symmetry, and a polytropic EOS with an adiabatic index $\Gamma=4/3$. The conservation equations result in a relativistic generalization of the Bernoulli equation: $\alpha\,h\,W=K$, where $K$ is a constant. Using that, along with the assumption of cold flow ahead of the shock (zero pressure), we find $K=c^{2}$, and with that compute the fluid field profiles ahead of the shock. Just ahead of the shock we apply the Rankine-Hugoniot jump conditions to obtain the values of the fluid field variables just below the shock, and with those we solve the GRHD equations again to obtain the profiles everywhere below the shock. For numerical reasons we cannot use a pre-shock pressure of zero, so instead we assume highly supersonic flow with a Mach number of 10, where the Mach number $M\equiv \left|v\right|/c_{s}$, where $c_{s}$ is the speed of sound.

\begin{figure}
\centering
\includegraphics[width=\textwidth]{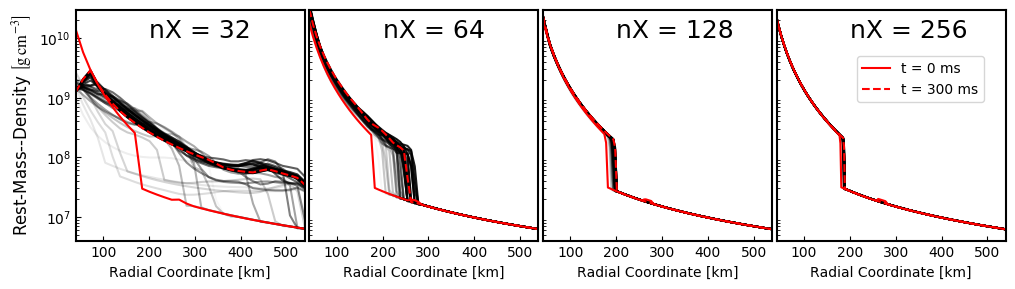}
\caption{This plot shows the primitive rest-mass density, $\rho$, throughout the evolution of a GR, 1D SAS with four different resolutions. The evolution is shown in shades of grey, with later times being darker. The initial and final states are shown with red solid and dashed lines, respectively. The perturbation is an overdense shell that passes through the shock, and with sufficient resolution the shock moves by only a few kilometers over the 300 ms interval, an interval that is several times longer than the typical timescale of the evolution of the instability, when it develops.}
\label{Fig:GR_SAS}
\end{figure}

We choose an accretion rate of $0.3\,\Msun/s$, and a spatial domain that extends from 40 km (the assumed radius of the PNS) to 540 km, three times the chosen initial shock radius of 180 km. On top of the density field we superpose a perturbation in the form of a 10\% overdense shell, which is accreted through the shock. This generates waves that propagate to the inner boundary, which are then reflected. It is known that with a non-relativistic treatment the stalled shockwave is stable to radial perturbations \cite{Blondin2003}. In \figref{Fig:GR_SAS}, we show results from runs using four different resolutions, and demonstrate that in order for the shock to maintain its initial position to an accuracy of $\mc{O}\left(10\%\right)$, using third-order methods with the HLL Riemann solver \cite{Harten1983} (see below), at least 128 elements are required; a coarser resolution leads to the shock deviating too far from its original position. We see from \figref{Fig:GR_SAS} that for a resolution greater than or equal to 128 (corresponding to a radial mesh size of about 3.9 km) the shock recedes by only a few kilometers over the 300 ms evolution, an encouraging result. All of these runs used $C_{\text{CFL}}=0.5$.


For the SAS runs we use the HLL Riemann solver. This is because of the known ``odd-even" decoupling phenomenon \cite{Quirk1994}, in which shocks aligned with the coordinate directions are artificially unstable with some Riemann solvers; e.g., HLLC. As shown in \cite{Quirk1994} the HLL solver is free from this issue. In the future we will incorporate a hybrid algorithm that uses the HLLC solver in smooth regions and switches to the HLL solver near shocks \cite{Muller2010}.

We find the outcomes of \cite{Blondin2003} are unchanged when using GRHD. In addition to the shock's stability to radial perturbations, it is also well known that the shock is \textit{un}stable to \textit{non}-radial perturbations \cite{Blondin2003}, the instability being known as the standing accretion shock instability (SASI). We investigate the GR version of the SASI in the next section.

\subsection{Relativistic 2D SAS Problem}

\begin{figure}
\centering
\includegraphics[width=\textwidth]{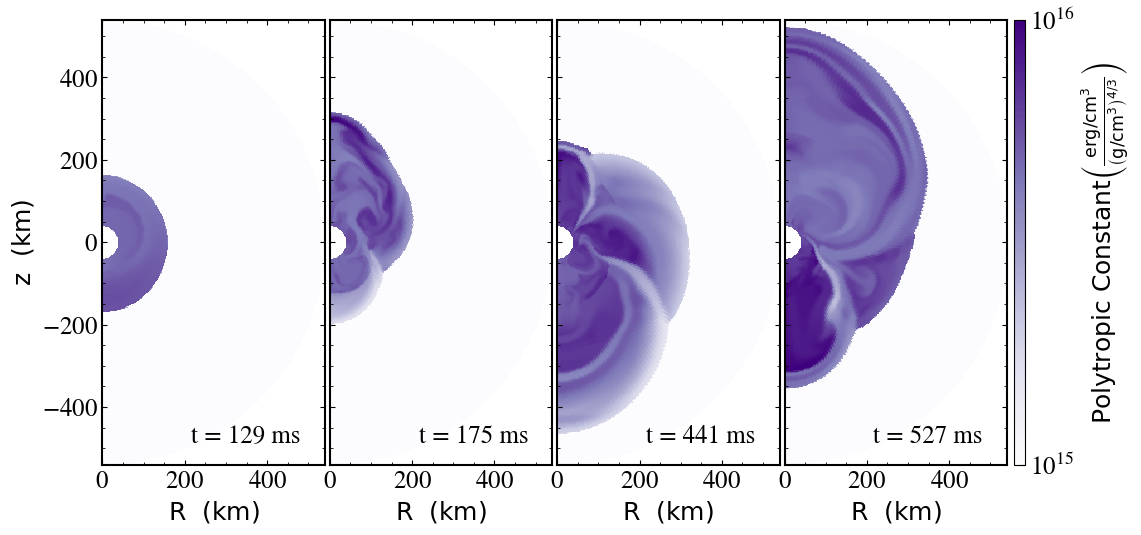}
\caption{Log plot of the polytropic constant (a proxy for entropy) of the 2D, GR SAS shown at four different times during the evolution of the system: 129 ms, when the shock is still fairly spherical; 175 ms, when the shock has sloshed toward the north pole; 441 ms, when the shock has sloshed back towards the south pole; and 527 ms, the last snapshot before the shock reaches the outer boundary.}
\label{Fig:GR_SASI}
\end{figure}

The initial conditions for this problem are a PNS radius of 40 km, a PNS mass of 2.8 $M_{\odot}$, an initial shock radius of $180$ km, and an accretion rate of 3.0 $M_{\odot}\,s^{-1}$ at the initial shock radius. These values are chosen to promote the GR effects, and are inspired from \cite{Walk2006}. The computational domain and adiabatic index are the same as for the 1D case. The form of the perturbation for this problem is designed to excite the $\ell=1$ mode (known to be the dominant mode of the SASI in 2D \cite{BlondinMezzacappa2006}), and therefore varies with $\cos\theta$. As this asymmetric shell passes through the shock, the shock is perturbed in both the radial and angular directions. This instability grows with time, developing into a sloshing motion characteristic of the SASI. Shown in \figref{Fig:GR_SASI} is the polytropic constant $K=p\,\rho^{-\gamma}$ (a proxy for entropy) at four times during the evolution: 129 ms, 175 ms, 441 ms, and 527 ms. The times were chosen to show the sloshing, and the last snapshot is the final snapshot before the shock reaches the outer boundary.

\begin{figure}
\centering
\includegraphics[width=0.8\textwidth]{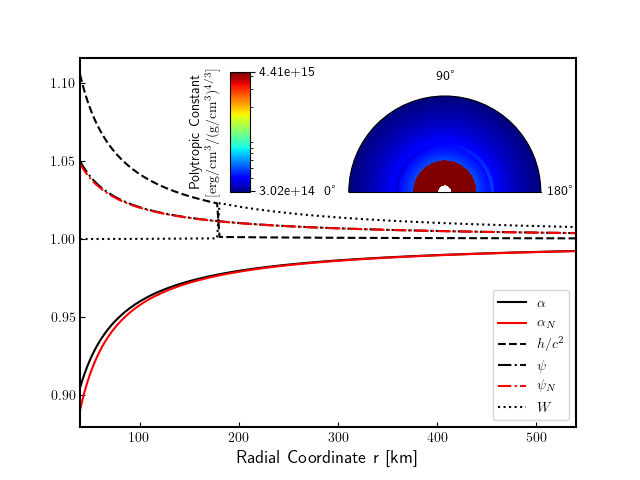}
\caption{Plotted against the radial coordinate $r$ are: the lapse function $\alpha$, the lapse function in the weak field limit $\alpha_{N}$, the specific enthalpy $h$ (normalized to the speed of light), the conformal factor $\psi$, the conformal factor in the weak field limit $\psi_{N}$, and the Lorentz factor $W$, all for the initial conditions. The Lorentz factor and the specific enthalpy are angle-averaged quantities, while (for this problem) the conformal factor and lapse function are intrinsically 1D and time-independent. Since $\alpha_{N}$ and $\psi_{N}$ deviate from unity by $\mathcal{O}\left(20\%\right)$, this problem is securely in the GR regime in terms of spacetime curvature, and the deviation of the specific enthalpy and the Lorentz factor from unity shows that this problem is GR in terms of the fluid velocities, both bulk and thermal. Inset: a 2D plot of the polytropic constant for the initial conditions. The perturbation is visible as a shell that varies in magnitude with the cosine of the polar angle. The polytropic constant of the shell is higher (darker color) than the ambient value near the north pole and lower (lighter color) than the ambient value near the south pole.}
\label{Fig:GR1D}
\end{figure}

The main physical difference between this test and that performed in \cite{Blondin2003} is: we treat the fluid as relativistic. The fluid could behave relativistically for several reasons. The bulk motion of the fluid could be comparable to the speed of light; this manifests as a Lorentz factor that deviates noticeably from unity. The internal fluid motions could be comparable to the speed of light; this manifests as a specific enthalpy that is not dominated by the rest-mass energy density of the fluid--i.e., $h/c^2>1$. The curvature of spacetime could be greater than what would be expected from Newtonian gravity; this manifests as a lapse function that differs from the Newtonian value, $\alpha_{N}=1+2\,\Phi/c^{2}$, where $\left|\Phi\right|/c^{2}\ll1$. These values are all shown in \figref{Fig:GR1D}, where we quantify the effects of GR for our setup. The Schwarzchild radius for our setup is $R_{s}=2GM/c^{2}\approx9$ km, about a quarter of our chosen PNS radius. Our choice of parameters was designed to test the code under GR conditions found after bounce in CCSNe. The deviation from the Newtonian case is best reflected in the plot of $\alpha_{N}$, where the deviation from unity reaches approximate 10\% at the inner boundary. In the Newtonian limit, $\left|\Phi\right|/c^{2}\ll1$, which is barely satisfied by the values we obtain here. With our new capability we intend to explore the SASI further by varying the values of parameters that promote relativistic effects expected over the range of post-bounce conditions obtained in detailed CCSN simulations.

\section{Summary/Future Work}
We have presented a solver for the GRHD equations under the conformally-flat approximation of GR using a DG method. Our solver has been tested against several challenging test problems in special and general relativistic regimes. The solver has incorporated parallel capabilities from \amrex, and we have successfully run problems using MPI. We find that the results from the test problems we run with \thornado\ agree well with published results (when available) and our scheme is able to maintain the steady state of the 1D SAS when the shock is perturbed by an overdense shell. We are also able to reproduce the SASI in 2D using GRHD, the first time the SASI has been studied in this way. Our results agree qualitatively with those in \cite{Blondin2003}. In a future publication, we will report on a more complete analysis of the development of the SASI in the GR case, along with detailed comparisons with the Newtonian case. These results are encouraging because the SASI plays an important role in CCSN simulations, and successfully modeling it in 1D and 2D is an auspicious start to modeling it in 3D.

Our next steps include implementing and running 3D problems---in particular a 3D GR-SASI---and investigating the effects of GRHD on the dynamics and evolution of the system. Although our method of slope-limiting works well, we are also looking into more sophisticated methods, such as borrowing from finite-volume WENO methods using sub-cell resolution, and/or $hp$-adaptivity; e.g., \cite{Dumbser2014,Fambri2018}. We also will be incorporating a realistic, tabular equation of state, GR gravity under the CFA, neutrino transport, as well as AMR within the \amrex\ framework, into \thornado. We will perform more in-depth timing studies and determine how best to optimize the code with respect to using multiple MPI ranks. We are also working to port \thornado\ to GPUs, which will greatly reduce the total run time for fully 3D CCSN simulations. The neutrino transport component has already been ported to GPUs \cite{Laiu2019} (in prep.), and we have begun the port of the hydrodynamics, with promising results.

\section*{Acknowledgements}
SJD, EE, and AM acknowledge support from the NSF Gravitational Physics Program (NSF-GP 1505933 and 1806692). This research made use of the software packages \texttt{AMReX}\footnote{https://amrex-codes.github.io/}, \texttt{Matplotlib} \cite{Hunter2007}, \texttt{NumPy} \cite{OliphantNumPy}, and \texttt{yt} \cite{Turk2011yt}. This work was conducted in part using the resources of the Advanced Computing Center for Research and Education at Vanderbilt University, Nashville, TN\footnote{https://www.vanderbilt.edu/accre/}. We acknowledge helpful interactions with Ann S. Almgren and Donald E. Wilcox with regard to interfacing \thornado\ with \amrex.

\section*{References}
\bibliography{bibfile}

\end{document}